\documentclass[twocolumn,superscriptaddress,amsmath,amssymb,prl]{revtex4-2}
\usepackage[pdftex,colorlinks=true,linkcolor=blue,citecolor=blue,urlcolor=blue]{hyperref}
\usepackage{graphicx,hyperref,tikz,scalerel}
\hypersetup{colorlinks=true, citecolor=blue, urlcolor=blue, linkcolor=blue} 
\renewcommand{\section}[1]{{\par\it #1.---}\ignorespaces}
\usetikzlibrary{svg.path}
\definecolor{orcidlogocol}{HTML}{A6CE39}
\tikzset{orcidlogo/.pic={
		\fill[orcidlogocol] svg{M256,128c0,70.7-57.3,128-128,128C57.3,256,0,198.7,0,128C0,57.3,57.3,0,128,0C198.7,0,256,57.3,256,128z};
		\fill[white] svg{M86.3,186.2H70.9V79.1h15.4v48.4V186.2z}
		svg{M108.9,79.1h41.6c39.6,0,57,28.3,57,53.6c0,27.5-21.5,53.6-56.8,53.6h-41.8V79.1z M124.3,172.4h24.5c34.9,0,42.9-26.5,42.9-39.7c0-21.5-13.7-39.7-43.7-39.7h-23.7V172.4z}
		svg{M88.7,56.8c0,5.5-4.5,10.1-10.1,10.1c-5.6,0-10.1-4.6-10.1-10.1c0-5.6,4.5-10.1,10.1-10.1C84.2,46.7,88.7,51.3,88.7,56.8z};}}
\newcommand\orcid[1]{\href{https://orcid.org/#1}{\mbox{\scalerel*{\begin{tikzpicture}[yscale=-1,transform shape]\pic{orcidlogo};\end{tikzpicture}}{|}}}}

\begin{document}
\title{Multiple topological corner states in the continuum of extended kagome lattice}
\author{Shun-Peng Zhang}
\affiliation{Key Laboratory of Quantum Theory and Applications of MoE \& Lanzhou Center for Theoretical Physics, Lanzhou University, Lanzhou 730000, China}
\affiliation{Gansu Provincial Research Center for Basic Disciplines of Quantum Physics
 \& Key Laboratory of Theoretical Physics of Gansu Province, Lanzhou University, Lanzhou 730000, China}
\author{Ming-Jian Gao\orcid{0000-0002-6128-8381}}
\affiliation{Key Laboratory of Quantum Theory and Applications of MoE \& Lanzhou Center for Theoretical Physics, Lanzhou University, Lanzhou 730000, China}
\affiliation{Gansu Provincial Research Center for Basic Disciplines of Quantum Physics
 \& Key Laboratory of Theoretical Physics of Gansu Province, Lanzhou University, Lanzhou 730000, China}
\author{Wei Jia}
\email{jiaw@lzu.edu.cn}
\affiliation{Key Laboratory of Quantum Theory and Applications of MoE \& Lanzhou Center for Theoretical Physics, Lanzhou University, Lanzhou 730000, China}
\affiliation{Gansu Provincial Research Center for Basic Disciplines of Quantum Physics
 \& Key Laboratory of Theoretical Physics of Gansu Province, Lanzhou University, Lanzhou 730000, China}
\author{Jun-Hong An\orcid{0000-0002-3475-0729}}
\email{anjhong@lzu.edu.cn}
\affiliation{Key Laboratory of Quantum Theory and Applications of MoE \& Lanzhou Center for Theoretical Physics, Lanzhou University, Lanzhou 730000, China}
\affiliation{Gansu Provincial Research Center for Basic Disciplines of Quantum Physics
 \& Key Laboratory of Theoretical Physics of Gansu Province, Lanzhou University, Lanzhou 730000, China}
\begin{abstract}
The kagome lattice is renowned for its exotic electronic properties, such as flat bands, Dirac points, and Van Hove singularities. These features have provided a fertile ground for exploring exotic quantum phenomena. Here, we discover that a breathing kagome lattice with long-range hoppings can host multiple zero-energy corner states, which emerge as topologically protected bound states in the continuum (BICs). This result demonstrates that additional hopping control can induce further non-trivial physics of the kagome lattice. Since the zero-energy corner states in the continuum are intertwined with a substantial number of zero-energy bulk states, we also develop a momentum-space topological characterization theory to precisely quantify the number of corner states, revealing a general bulk-corner correspondence. Furthermore, we uncover three distinct types of topological phase transitions (TPTs) for the BICs driven by shifts in the spatial localization of zero-energy bulk and/or edge states. These TPTs are exactly captured by our characterization theory. This work provides deep insights into the topological physics of the kagome lattice and broadens the understanding of its electronic properties.
\end{abstract}
\maketitle

{\it\color{blue}Introduction.}---The kagome lattice~\cite{10.1143/ptp/6.3.306} has attracted a persistent attention of condensed matter physics due to its unique electronic properties, such as flat bands~\cite{li2018realization,PhysRevB.99.125131,kang2020topological,PhysRevLett.126.196403,Yin2022}, Dirac points~\cite{ye2018massive,kang2020dirac,li2021dirac}, and Van Hove singularities~\cite{kang2022twofold,hu2022rich,jiang2024van,li2024intertwined}.
Since this two-dimensional ($2$D) lattice is constructed by the corner-sharing triangles arranged in a tessellation of hexagons, it not only models the molecular structures of many crystalline materials~\cite{Pereiro2014,Huang2017,PhysRevB.101.100405,Ghimire2020,PhysRevB.104.L041103,PhysRevB.104.L161115,Yu2021,Zhao2021,Xu2022,Neupert2022,Chowdhury2023,PhysRevB.110.085112}, but also serves as a fertile ground for exploring exotic quantum phenomena, including topological phases~\cite{PhysRevB.80.113102,Ni_2017,PhysRevB.99.165141,PhysRevB.102.245133,PhysRevResearch.2.022043,PhysRevB.103.195105,PhysRevB.104.144422,ZHAO2022105360,you2023topologicalchiralkagomelattice,Mojarro_2024}, quantum spin liquids~\cite{Han2012}, and unconventional superconductivity~\cite{Zhao2021,Neupert2022}. To date, the exploration and understanding of novel physical phenomena in kagome lattice have remained a significant and compelling area of research.

Recently, it is found that the kagome lattice with different strengths of nearest-neighbor hopping, i.e., breathing kagome lattice, can support second-order topological insulators (SOTIs), hosting single zero-energy topological state at each corner~\cite{PhysRevLett.120.026801,PhysRevB.97.241405,PhysRevB.99.085406,Sil_2020,PhysRevB.101.094107,PhysRevB.101.195143,PhysRevB.105.085411,PhysRevB.106.085420,PhysRevE.108.024112,Yatsugi2023,PhysRevA.110.013309,PhysRevB.109.165406,PhysRevB.110.104307}. This discovery significantly broadens the understanding of the interplay between geometry and topology in the kagome lattice. Notably, adding long-range hoppings in square lattice systems has been shown to enrich their topological properties~\cite{PhysRevLett.128.127601,PhysRevLett.131.157201,PhysRevApplied.20.064042,PhysRevB.110.L201117}. A natural question arises: does the kagome lattice host even more novel topological physics when its geometric structure is changed beyond conventional nearest-neighbor hopping framework? On the other hand, a recent breakthrough has demonstrated that topological corner states can be embedded in the continuum of a 2D Su–Schrieffer–Heeger lattice~\cite{PhysRevLett.118.076803}, revealing an intriguing connection between the SOTIs and bound states in the continuum (BICs)~\cite{hsu2016bound}. Despite being intertwined with numerous zero-energy bulk states, these topological BICs become fully localized at the corners when the necessary symmetries are present~\cite{PhysRevB.101.161116}. Traditionally, the BICs are unique solutions to wave equations, being discrete and spatially bounded, but at the same energy as a continuum of states propagating to infinity~\cite{PhysRevLett.113.257401}. Due to their novel physics and potential applications~\cite{meier1999laser,noda2001polarization,yanik2011seeing,PhysRevB.89.165111,hirose2014watt}, they have been theoretically studied~\cite{PhysRevB.39.5476,PhysRevLett.90.013001,PhysRevLett.102.167404} and experimentally realized in various systems, including waveguide array~\cite{PhysRevLett.107.183901,PhysRevA.93.062122,PhysRevLett.125.213901}, photonic system~\cite{PhysRevLett.100.183902,hu2021nonlinear}, and topological circuits~\cite{PhysRevLett.132.046601}. However, it remains unexplored in the kagome systems. This raises another question: can the kagome lattice host topological BICs when its geometric structure is tuned?

In this Letter, we address the above two significant questions. We first demonstrate that a breathing kagome lattice with long-range hoppings can host multiple zero-energy corner states, which emerge as the topological BICs protected by three-fold rotation and mirror symmetries. By further employing two types of special discrete momentum points located in the Brillouin zone (BZ), we define a momentum-space  topological invariant to precisely characterize the corner states in the continuum. This result is out of one's expectation and reveals a general bulk-corner correspondence of kagome systems. Furthermore, when the hopping parameters of the lattice are tuned, we discovery three distinct types of topological phase transitions (TPTs) in the BICs, which are driven by shifts in the spatial localization of bulk and/or edge zero-energy states. By observing the behaviors of two types of momentum points, these TPTs are exactly captured through our theory. This work promotes the studies of topological physics in the kagome lattice beyond the conventional hopping framework and provides an insight into its electronic properties.

\begin{figure}[!t]
\centering
\includegraphics[width=1\columnwidth,trim=0cm 0 0cm 0,clip=false]{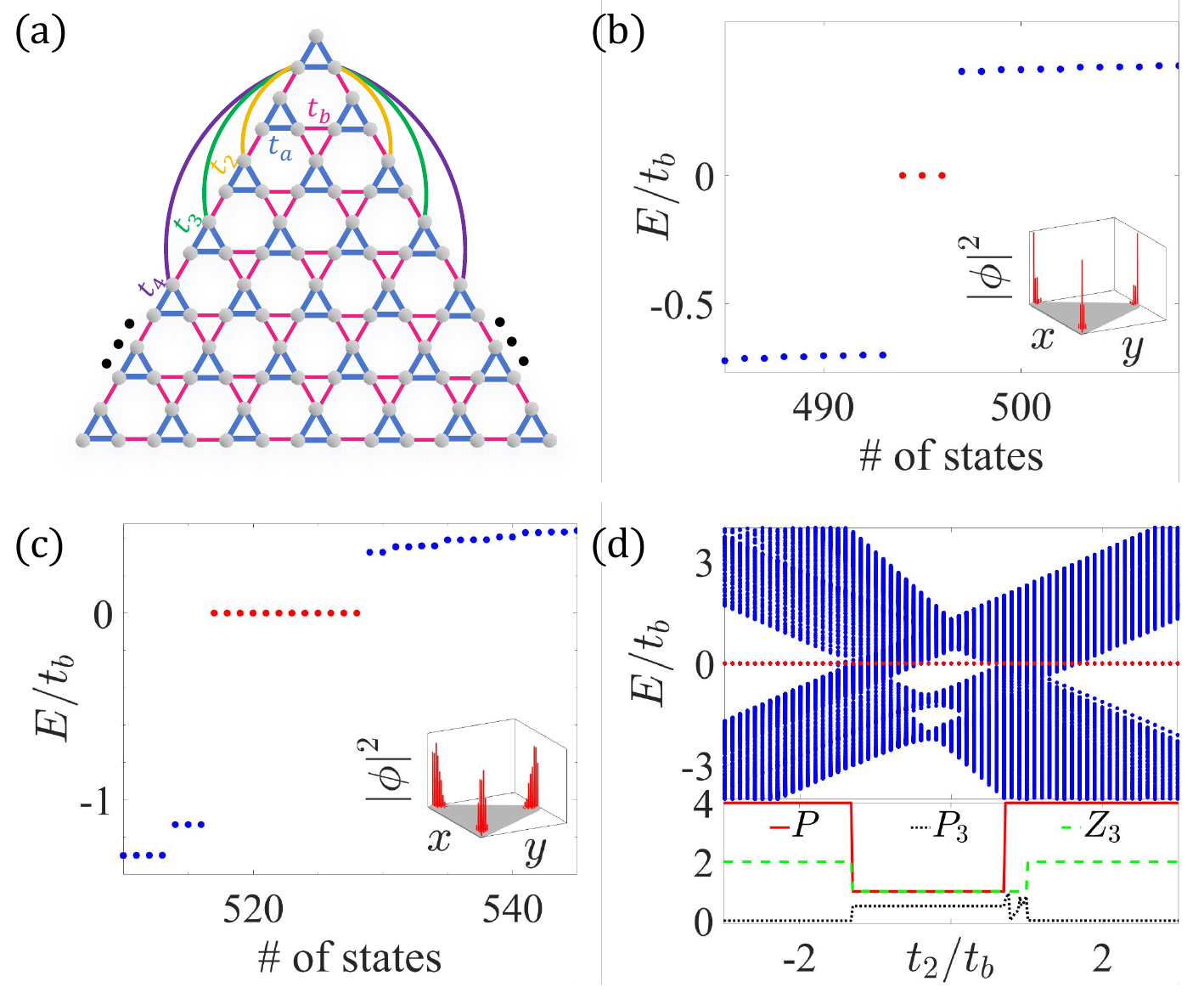}
\caption{(a) Schematic diagram of the breathing kagome lattice with long-range hoppings. Zero-energy topological states and their probability distributions for (b) $t_2=0$ and (c) $t_2=2t_b$. (d) Energy spectrum under the $x$- and $y$-direction open boundary conditions and the topological invariants $P_3$, $Z_3$, and $P$. We use $N_x=N_y=30$, $t_a=0.3t_b$, and $ t_{m>2}=0$.}
\label{fig:1}
\end{figure}

{\it\color{blue}Model.}---Our starting point is a $2$D spinless fermion model on the breathing kagome lattice with long-range hoppings, as shown in Fig.~{\ref{fig:1}}(a). The corresponding momentum-space Hamiltonian reads
\begin{equation}
\begin{split}
H(\mathbf{k}) =&-\left(
\begin{array}{ccc}
0 & h_{12} & h_{13} \\
h_{12}^{\ast } & 0 & h_{23} \\
h_{13}^{\ast } & h_{23}^{\ast } & 0%
\end{array}%
\right),
\end{split}
\label{eq:Hamiltonian}
\end{equation}
with $h_{12}=t_{a}+t_{b}e^{-i\mathbf{k\cdot a}_{3}}+\sum^M_{m=2}t_{m}e^{-im\mathbf{k\cdot a}_{3}}$, $h_{13}=t_{a}+t_{b}e^{-i\mathbf{k\cdot a}_{2}}+\sum^M_{m=2}t_{m}e^{-im\mathbf{k\cdot a}_{2}}$, and $h_{23}=t_{a}+t_{b}e^{-i\mathbf{k\cdot a}_{1}}+\sum^M_{m=2}t_{m}e^{-im\mathbf{k\cdot a}_{1}}$. The lattice vectors are $\mathbf{a}_{1} =\left( 1/2,-\sqrt{3}/2\right)$, $\mathbf{a}_{2} =\left( 1,0\right)$, and $
\mathbf{a}_{3} =\left( 1/2,\sqrt{3}/2\right)$. The momentum is $\mathbf{k}=(k_x,k_y)$. $t_a$ and $t_b$ are the nearest-neighbor hoppings within a unit cell and between adjacent unit cells, respectively. $t_m$ with $m=2,3,\cdots,M$ are the hopping rates between the $m$th nearest-neighbor unit cells. Each unit cell of the model has three lattice sites and thus its bulk Hamiltonian hosts three energy bands. However, two bands reduce to a two-fold degenerate energy level at the high-symmetry point $\boldsymbol{\Gamma}=(0,0)$ and the system only has one band gap. It makes the system exhibit either an insulator or a metal phase determined by whether the Fermi energy $E_F=0$ is located at the middle/lower or the upper of the bulk energy gap.

\begin{figure}[!b]
\centering
\includegraphics[width=1\columnwidth,trim=0cm 0 0cm 0cm,clip=false]{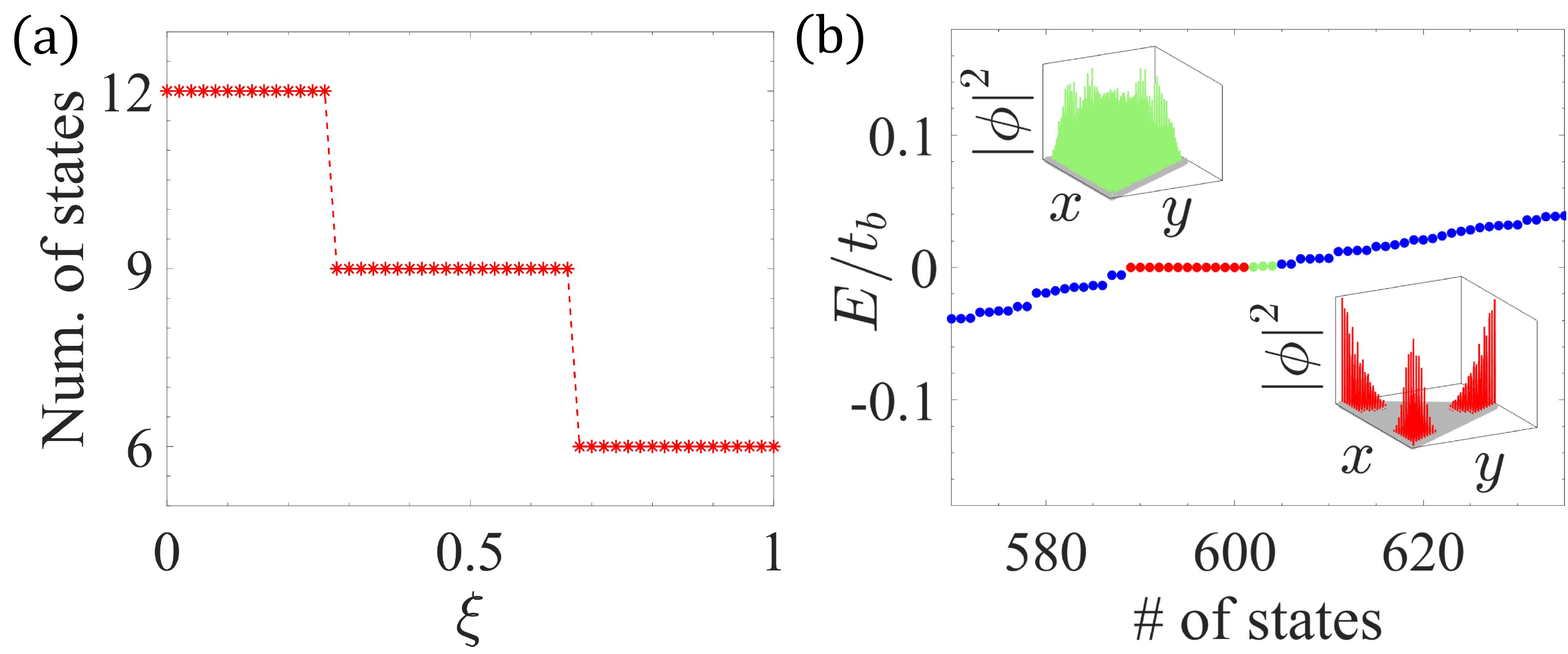}
\caption{(a) Number of zero-energy corner states in the continuum in different disorder strength $\xi$. (b) Distribution of zero-energy states in the continuum for the weak disorder with $\xi=0.02$, where there are $12$ ($3$) zero-energy states in the corners (bulk). We use $t_a=0.3t_b$, $t_2=0.9t_b$, and $t_{m>2}=0$.}
\label{fig:3}
\end{figure}

The long-range hoppings $t_m$ induce the system to host rich topological states. In the absence of $t_m$, this model describes a traditional breathing kagome lattice and renders a SOTI phase~\cite{PhysRevLett.120.026801}, which hosts an in-gap zero-energy state at each corner [see Fig.~\ref{fig:1}(b)]. Being protected by three-fold rotation symmetry with $C_3\mathcal{H}(\mathbf{k})C_3^{-1}=\mathcal{H}(R_3\mathbf{k})$, mirror symmetry with $M_x\mathcal{H}(k_x,k_y)M_x^{-1}=\mathcal{H}(-k_x,k_y)$, time-reversal symmetry with ${\Theta}{\cal H}\left( \mathbf{k}\right){\Theta^{-1}} ={\cal H}\left( -\mathbf{k}\right)$, and  generalized chiral symmetry with $\mathcal{H}(\mathbf{k})+\Gamma_3\mathcal{H}(\mathbf{k})\Gamma_3^{-1}+\Gamma_3^2\mathcal{H}(\mathbf{k})\Gamma_3^{-2}=0$~\cite{PhysRevB.103.205306,PhysRevB.111.085150}, where
\begin{equation}
\begin{split}
&{{C}_3}=\left(
\begin{array}{ccc}
0 & 1 & 0 \\
0 & 0 & 1 \\
1 & 0 & 0%
\end{array}
\right),~~
{{ M}_x}=\left(
\begin{array}{ccc}
0 & 0 & 1 \\
0 & 1 & 0 \\
1 & 0 & 0%
\end{array}
\right),\\
&{{ \Theta}}=\left(
\begin{array}{ccc}
1 & 0 & 0 \\
0 & 1 & 0 \\
0 & 0 & 1%
\end{array}
\right)\mathcal{K},~~
{{ \Gamma}_3}=\left(
\begin{array}{ccc}
1 & 0 & 0 \\
0 & e^{i2\pi/3} & 0 \\
1 & 0 & e^{-i2\pi/3}%
\end{array}
\right),
\end{split}
\label{eq:symmetry}
\end{equation} 
and $R_3$ denotes to anti-clockwisely rotate $\mathbf{k}$ by $2\pi/3$, such topological corner state is characterized by a $\mathbb{Z}_2$ bulk polarization $P_3=1/2$~\cite{PhysRevLett.120.026801,PhysRevA.110.013309,doi:10.1126/science.aah6442,PhysRevB.98.201402,PhysRevB.96.245115} or a $\mathbb{Z}_3$ Berry phase $Z_3=1$ after being divided by $2\pi/3$ for convenience~\cite{PhysRevLett.120.247202,PhysRevResearch.2.022028,PhysRevB.101.094107,PhysRevB.101.195143,PhysRevResearch.2.012009}. The presence of $t_m$ terms, preserving the above symmetries, induces the emergence of multiple zero-energy states at each corner, which are located not only in the bulk energy gap but also in the continuum. In the former case, $E_F$ is in the middle of bulk energy gap and the system is a SOTI with multiple in-gap zero-energy corner states [see Fig.~{\ref{fig:1}}(c)]. In the latter case, $E_F$ is in the lower or upper of bulk energy gap and the system hosts the topologically protected BICs [see Fig.~\ref{fig:1}(d)]. To verify the topological-protection feature of these zero-energy corner states in the continuum, we add a disorder $\eta\xi$ on $t_2$ so that the hopping rate becomes $t_2+\eta\xi$, where $\eta$ is a random number in the regime of $[-1, 1]$ and $\xi$ is the disorder strength. It is observed that the zero-energy corner states in the continuum are robust to the weak disorder (see Fig.~\ref{fig:3}), although this disorder breaks $C_3$ and $M_x$ symmetries. The reason is that $\Gamma_3$ and $\Theta$ are still preserved, resulting in the well-protected zero-energy corner states~\cite{PhysRevLett.128.127601}. However, it is noted that such zero-energy corner states cannot be characterized by $P_3$ and $Z_3$. We next propose a generic momentum-space topological invariant to characterize these zero-energy corner states.

\begin{figure}[!t]
\centering
\includegraphics[width=1\columnwidth,trim=0cm 0 0cm 0cm,clip=false]{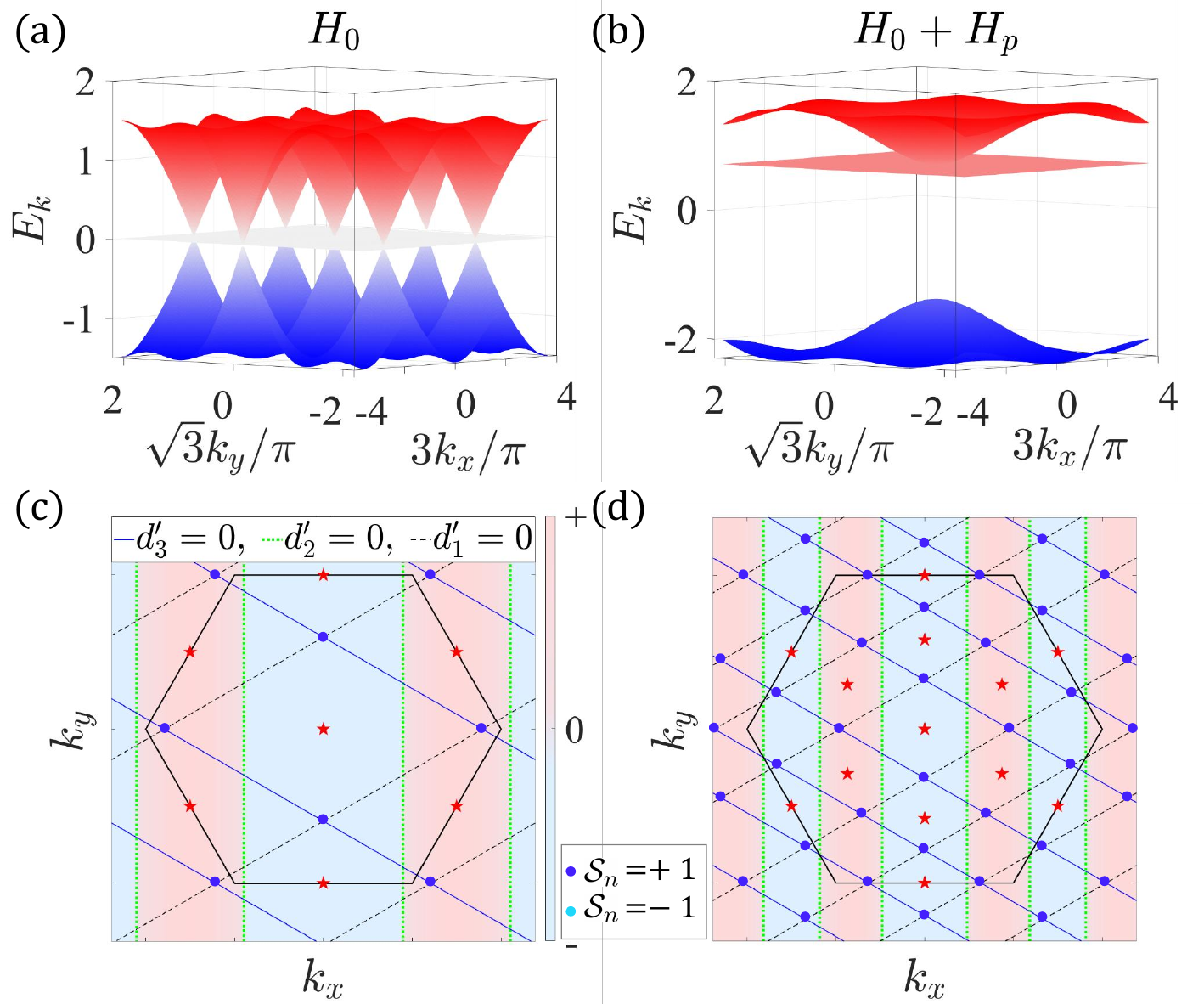}
\caption{(a) Energy bands of ${H}_0(\mathbf{k})$. It hosts four Dirac points. (b) Energy bands of ${H}(\mathbf{k})$. The energy gap is open. (c)-(d) Topological characterization for (c) $t_2=0$ and (d) $t_2=2.0t_b$. The red stars denote $\boldsymbol{\varrho}$ points determined by $d_{1,2,3}(\mathbf{k})=0$. The blue solid circles denote $\boldsymbol{\vartheta}$ points determined by $d'_{1,3}(\mathbf{k})=0$. The light-red and light-blue regions denote $d'_2(\mathbf{k})>0$ and $d'_2(\mathbf{k})<0$, respectively. The black hexagons denote the boundaries of the BZ. We use $t_a=0.3t_b$ and $t_{m>2}=0$.
}
\label{fig:2}
\end{figure}

{\it\color{blue}Topological characterization.}---The topological invariant, say $P$, is defined by the features of two types of discrete momentum points in the BZ. We first separate the bulk Hamiltonian ${H}(\mathbf{k})$ into two parts, i.e.,
\begin{equation}
{H}(\mathbf{k})={H}_0(\mathbf{k})+{H}_p(\mathbf{k}),
\end{equation}
with ${H}_0(\mathbf{k})=\mathbf{d}(\mathbf{k})\cdot\mathbf{S}$, ${H}_p(\mathbf{k})=\mathbf{d}'(\mathbf{k})\cdot\mathbf{S}'$, $\mathbf{S}=(\lambda_7,\lambda_5,\lambda_2)$, and $\mathbf{S}'=(\lambda_6,\lambda_4,\lambda_1)$. Here $\lambda_i$ are six ones of the eight Gell-Mann matrices~\cite{Gell-mann2000}. Then, we obtain $\mathbf{d}(\mathbf{k})$ and $\mathbf{d}'(\mathbf{k})$ as follows:
\begin{equation}
\begin{split}
&{d_j}(\mathbf{k})=-t_b\sin(\mathbf{k\cdot a}_{j})-\sum_{m=2}^Mt_m\sin(m\mathbf{k\cdot a}_{j}),\\
&{d'_j}(\mathbf{k})=-t_a-t_b\cos(\mathbf{k\cdot a}_{j})-\sum_{m=2}^Mt_m\cos(m\mathbf{k\cdot a}_{j}),
\end{split}
\end{equation}
with $j=1,2,3$. It is clear that ${H}_0(\mathbf{k})$ describes a Dirac semimetal and hosts several Dirac points, as shown in Fig.~\ref{fig:2}(a). These nodes define the first type of momentum points, $\boldsymbol{\varrho}\equiv\{\mathbf{k}\in\text{BZ}|d_j(\mathbf{k})=0,j=1,2,3\}$. The addition of ${H}_p(\mathbf{k})$ in $H_0(\mathbf{k})$ opens the energy gap of these Dirac points, as shown in Fig.~\ref{fig:2}(b). This makes the system be described by $H(\mathbf{k})$ and emerge zero-energy corner states. We then take any two components of ${\bf d}'({\bf k})$, say $d'_{\alpha}(\mathbf{k})$ and $d'_\beta(\mathbf{k})$, to define the second type of momentum points as $\boldsymbol{\vartheta}\equiv\{\mathbf{k}\in\text{BZ}|d'_\alpha(\mathbf{k})=d'_\beta(\mathbf{k})=0\}$. Accordingly, the remaining component $d'_\gamma(\mathbf{k})$ has nonzero value at both $\boldsymbol{\varrho}$ and $\boldsymbol{\vartheta}$ points due to the existence of the bulk energy gap, as shown in Fig.~\ref{fig:2}(c). It is noted that the number of $\boldsymbol{\varrho}$ or $\boldsymbol{\vartheta}$ points is always even, as the system is protected by $C_3$ and $M_x$ symmetries.

We next define the topological invariant $P$ by the features of $\boldsymbol{\varrho}$ and $\boldsymbol{\vartheta}$ points. Physically, the topological zero-energy corner states in 2D systems are induced by opening the energy gap of its $1$D edge states and producing mass domain walls at the cross point of the two involved edges~\cite{doi:10.1126/science.aah6442,PhysRevLett.121.096803,PhysRevB.96.245115,PhysRevLett.118.076803,doi:10.1126/sciadv.aat0346,PhysRevB.97.205136,PhysRevX.9.011012,PhysRevLett.123.177001,PhysRevLett.122.236401,https://doi.org/10.1002/pssb.202000090,PhysRevB.105.L041105}. Thus, these corner states should inherit the topology of the edge states~\cite{PhysRevResearch.5.L022032,PhysRevB.106.245105,LI20211502}.
Based on this fundamental idea, we find that the topology of edge states of this extended kagome model is characterized by a $\mathbb{Z}_2$ topological index, $P_1=\text{sgn}(\mathcal{N}_+\mathcal{N}_-)$, where $\mathcal{N}_+$ and $\mathcal{N}_-$ are the number of $\boldsymbol{\varrho}$ points in the regions of ${d'_\gamma}(\boldsymbol{\varrho})>0$ and ${d'_\gamma}(\boldsymbol{\varrho})<0$, respectively. When $P_1=0$, the edge states are absent. Thus, the system is topologically trivial in second order. When $P_1=1$, the edge states possessing different signs of $d'_\gamma({\boldsymbol \rho})$ exist in both edges. It causes the formation of the topological zero-energy corner states at the cross points of the two edges. Inspired by Ref.~\cite{PhysRevResearch.2.043136}, the two edge Hamiltonians of Eq.~\eqref{eq:Hamiltonian} are calculated; see more details in the Supplementary Material~\cite{Supplementary_Materials}. We discover that each edge is described by a 1D Su-Schrieffer-Heeger model with long-range hopping, in which the winding number of each edge Hamiltonian can be rigorously defined. The product of two winding numbers in the two edges gives the number of zero-energy corner state. It is further rewritten as $P_2=\sum_n{\cal S}_n/4$, where ${\cal S}_n={\cal C}_n{\cal P}_n$ defines the polarized topological charges at $\boldsymbol{\vartheta}$ points~\cite{PhysRevB.110.L201117}. Here the $n$th topological charge and its polarization are determined by
\begin{equation}
\begin{split}
{\cal C}_n=\text{sgn}[\frac{\partial d'_{\alpha}(\boldsymbol{\vartheta})}{\partial k_\alpha}\frac{\partial d'_{\beta}(\boldsymbol{\vartheta})}{\partial k_\beta}],~
{\cal P}_n=\text{sgn}[d_{\alpha}(\boldsymbol{\vartheta})d_{\beta}(\boldsymbol{\vartheta})],
\label{eq:C_n}
\end{split}
\end{equation}
respectively. Note that the partial differentials are calculated for the newly defined momentum $k_j=\mathbf{k}\cdot \mathbf{a}_j$ with $j=1,2,3$. Finally, we obtain the $\mathbb{Z}$ topological invariant
\begin{equation}
{P}=P_1P_2=\frac{1}{4}\text{sgn}\left(\mathcal{N}_+\mathcal{N}_-\right)\sum_{n}{\cal S}_{n}
\label{eq:topological invariants}
\end{equation}
to characterize the zero-energy corner states, which are not only located in the bulk gap but also in the continuum. This elegant result reveals a general bulk-corner correspondence of the kagome systems. In addition, it is worth mentioning that we have exactly proved Eq.~\eqref{eq:topological invariants} in the Supplementary Material~\cite{Supplementary_Materials}.

\begin{figure*}[!ht]
\centering
\includegraphics[width=2.0\columnwidth,trim=0.65cm 0cm 0cm 0cm,clip=false]{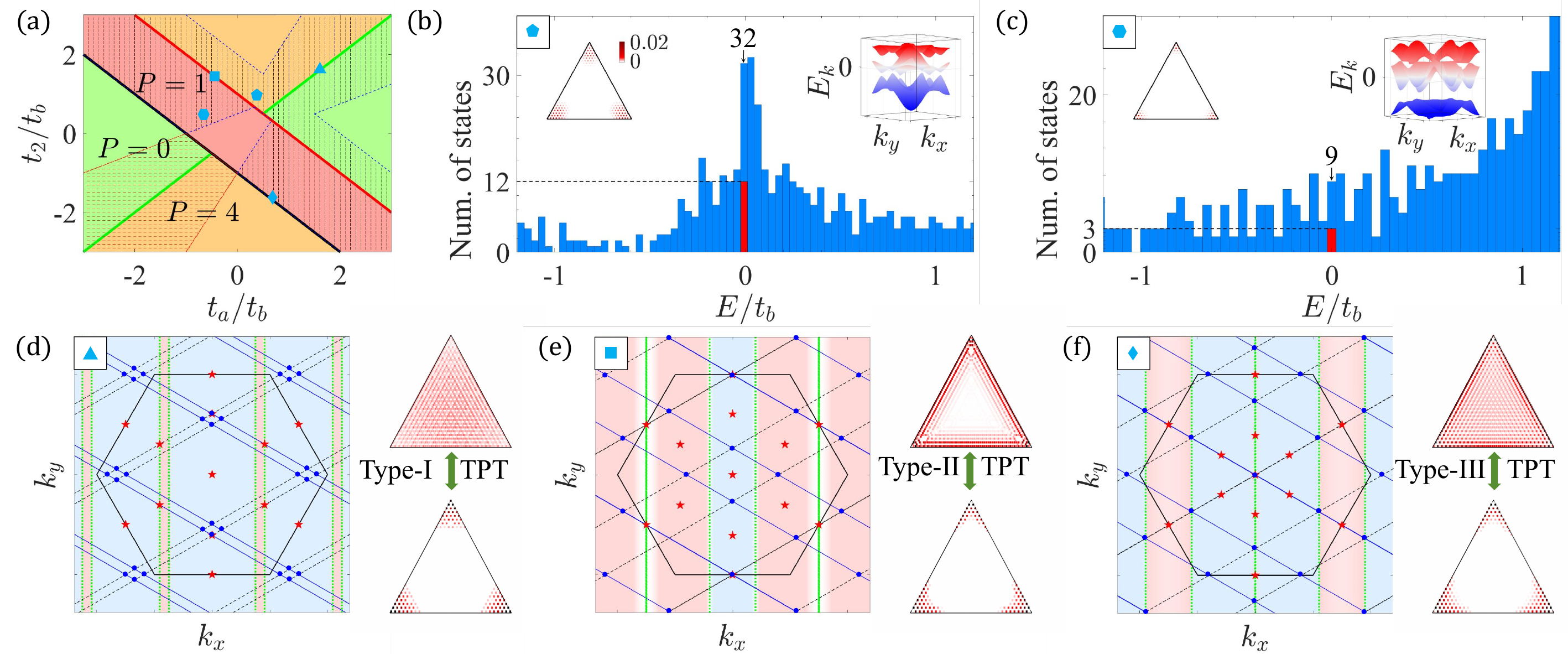}
\caption{(a) Topological phase diagram determined by $P$. The shadow black vertical dashed (red horizontal dashed) lines mark the regions where $E_F$ is located at the upper (lower) of bulk energy gap. Statistics diagram of the eigenstates of $H(\mathbf{k})$ under the $x$- and $y$-direction open boundary condition when (b) $t_a=0.3t_b$ and $t_2=0.9t_b$ and (c) $t_a=-0.5t_b$ and $t_2=0.7t_b$, respectively. These corner states marked by red color are in the continuum and characterized by $P=4$ and $P=1$. The insets show the distribution of zero-energy states and the momentum-space bulk energy bands, where $E_F$ is located at the upper of bulk energy gap, implying the metal phases. Different types of TPTs with (d) $t_a=1.52t_b$ and $t_2=1.52t_b$, (e) $t_a=-0.52t_b$ and $t_2=1.52t_b$, and (f) $t_a=0.52t_b$ and $t_2=-1.52t_b$ are directly identified by the behaviours of $\boldsymbol{\varrho}$ and $\boldsymbol{\vartheta}$ points.}
\label{fig:4}
\end{figure*}

{\it\color{blue} Realistic applications.}---The above topological characterization has theoretical simplicity and broad applicability. To further confirm these nontrivial properties, we first consider the case without long-range hoppings by taking $(\alpha,\beta,\gamma)=(3,1,2)$, i.e., the traditional breathing kagome model. It is observed from the red stars of Fig.~\ref{fig:2}(c) that the system has four $\boldsymbol{\varrho}$ points in the BZ. Only two $\boldsymbol{\varrho}$ points are located at the region of ${d'_2}(\boldsymbol{\varrho})<0$ and give $\mathcal{N}_{-}=2$, while the remaining two $\boldsymbol{\varrho}$ points exist in the regions of ${d'_2}(\boldsymbol{\varrho})>0$ and give $\mathcal{N}_{+}=2$. Meanwhile, we observe four $\boldsymbol{\vartheta}$ points in the BZ, hosting the polarized topological charges $\mathcal{S}_{1,2,3,4}=1$; see the blue solid circles of Fig.~\ref{fig:2}(c). We immediately identify $P=1$ for this case, which implies that there is only single zero-energy state at each corner, as shown in Fig.~\ref{fig:1}(b).

Next, we consider the extended kagome model where the system has the second nearest-neighbor hoppings. A similar method is used to determine $\boldsymbol{\varrho}$ and $\boldsymbol{\vartheta}$ points, as shown in Fig.~{\ref{fig:2}}(d). Then, the topological invariant is given by $P=4$, implying that there are four zero-energy states at each corner, as confirmed by Fig.~\ref{fig:1}(c). Furthermore, we calculate the phase diagram of this case, as shown in Fig.~{\ref{fig:4}}(a). It is observed that the system presents the normal insulators or metals when $P=0$. When $P\neq 0$, the phase regions with and without shadow lines are the SOTIs/metals with topologically protected BICs and the SOTIs with in-gap topological corner states, respectively. Our characterization theory can exactly describe the number of zero-energy corner states in the continuum [see Figs.~\ref{fig:4}(b) and \ref{fig:4}(c)]. We also consider the case with the long-range hoppings among the second and third nearest-neighbor unit cells. The result shows that the corner states characterized by $P=1$, $4$, and $9$ are still topologically protected BICs and emerge in a broad range of hopping parameters~\cite{Supplementary_Materials}.

{\it\color{blue} Topological phase transitions of BICs.}---We next show three distinct types of TPTs for the BICs driven by shifts in the spatial localization of zero-energy bulk and/or edge states. This mechanism captures all TPTs in the extended kagome system, even if the BICs emerge in the metal phase where no band energy gap exists. Specifically, when the parameters of system are adjusted, certain bulk-localized and/or edge-localized (corner-localized) zero-energy states can transform into corner-localized (bulk-localized and/or edge-localized) zero-energy states, thereby increasing (decreasing) the number of topological corner states. As shown in Fig.~\ref{fig:4}(a), these transformations exhibit three distinct types of TPTs for topological BICs: Type-I TPT marked by green phase boundary, where corner-localized states transition into bulk-localized states, see the inset of Fig.~{\ref{fig:4}}(d); Type-II TPT marked by red phase boundary, where corner-localized states transition into edge-localized states, see the inset of Fig.~{\ref{fig:4}}(e); Type-III TPT marked by black phase boundary, where corner-localized states transition into a coexistence of bulk- and edge-localized states, see the inset of Fig.~{\ref{fig:4}}(f). 

The above three nontrivial TPTs can be directly identified by our topological characterization theory. For the type-I TPT, we observe that certain $\boldsymbol{\vartheta}$ points within the BZ intersect with $\boldsymbol{\varrho}$ points but do not coincide with the momenta where $d'_\gamma(\mathbf{k})=0$, as shown in Fig.~{\ref{fig:4}}(d). This actually indicates that three bands touch; however, the touching points between the upper and middle bands differ from those between the middle and lower bands. For the type-II TPT, we find that $\boldsymbol{\varrho}$ points intersect with specific $\boldsymbol{\vartheta}$ points and overlap with the region where $d'_\gamma(\mathbf{k})=0$ on the boundary of the BZ, as shown in Fig.~\ref{fig:4}(e). This implies that only two of the three bands touch. For the type-III TPT, we observe that $\boldsymbol{\varrho}$ points, $\boldsymbol{\vartheta}$ points and the momentum of  $d'_\gamma(\mathbf{k})=0$ intersect at $\boldsymbol{\Gamma}$ point, as shown in Fig.~{\ref{fig:4}}(f). This corresponds to the case where three bands touch simultaneously at $\boldsymbol{\Gamma}$ point. These results further demonstrate that the momentum space information can essentially uncover the physical mechanism responsible for the localized position changes of the zero-energy states.

{\it\color{blue}Discussion and Conclusion.}---In the extended kagome systems described by Eq.~\eqref{eq:Hamiltonian}, the long-range hopping terms are set to span an even number of lattice sites, which ensures that this Hamiltonian can be exactly projected into its $1$D boundary. This hopping feature is crucial for the validity of our topological invariant $P$, which relies on the existence of well-defined boundary states. For the longer-range hopping terms which do not obey the previous constraint, this case results in delocalized or hybridized boundary modes and then undermines the applicability of the topological characterization framework~\cite{Supplementary_Materials}. Although our model focuses on a class of specific long-range hopping terms, such configurations are not limited to theoretical constructs and can be implemented in both natural and artificial systems. In realistic materials, the kagome metals, like CoSn and FeSn, provide a promising platform, where the effective hopping amplitudes can be modulated by external strain or chemical substitution~\cite{PhysRevB.79.214502,PhysRevB.104.L060405,PhysRevB.105.085138,PhysRevB.107.205130,PhysRevB.109.L201106}. In engineered systems, superconducting qubit arrays with tunable couplers allow for precise control over long-range correlations~\cite{PRXQuantum.4.030313,PhysRevB.110.054511}. In addition, topological circuits offer a highly tunable environment in which spatially varying hopping profiles can be implemented through the design of LC networks~\cite{PhysRevResearch.2.022028}.

In summary, we have investigated the topological phases of a breathing kagome model with long-range hoppings, which hosts multiple zero-energy corner states embedded within the continuum. By analyzing the features of two types of discrete momentum points in the Brillouin zone, we have established a momentum-space topological characterization theory that precisely determines these corner states in the continuum. It reveals a general bulk-corner correspondence. Furthermore, we have identified three distinct types of topological phase transitions for the topological corner states in the continuum, all of which are effectively captured by our theoretical framework. This work not only advances the understanding of topological physics in kagome systems but also offers valuable insights for exploring their broader electronic properties.

{\it\color{blue}Acknowledgements.}---This work is supported by the National Natural Science Foundation of China (Grant No. 12404318, No. 124B2090, No. 12275109, and No. 12247101),  the Innovation Program for Quantum Science and Technology
of China (Grant No. 2023ZD0300904), the Fundamental Research Funds for the Central Universities (Grant No. lzujbky-2024-jdzx06), the Natural Science Foundation of Gansu Province (No. 22JR5RA389), and the ‘111 Center’ under Grant No. B20063.

\pagebreak
\clearpage
\onecolumngrid
\flushbottom
\begin{center}
\textbf{\large Supplementary Material for ``Multiple topological corner states in the continuum of extended kagome lattice"}
\end{center}
\setcounter{equation}{0}
\setcounter{figure}{0}
\setcounter{table}{0}
\makeatletter
\renewcommand{\theequation}{S\arabic{equation}}
\renewcommand{\thefigure}{S\arabic{figure}}
\renewcommand{\bibnumfmt}[1]{[S#1]}
\renewcommand{\citenumfont}[1]{S#1}

In this Supplementary Material, we provide the details of the main text. In Sec.~I, we give the exact proof of our topological characterization theory. In Sec.~II, we show more numerical results of the extended kagome model.

\subsection{I. Exact proof of topological characterization theory}\label{one}

\subsubsection{a. Edge Hamiltonians of extended SSH model}

Inspired by Ref.~\cite{s-PhysRevResearch.2.043136}, we first introduce the forward and backward shift operators of a lattice system. They are defined as 
\begin{equation}
\delta f_j=f_{j+1},~~\delta^{*} f_j=f_{j-1},
\label{forward_backward_shift_operator}
\end{equation}
where $f_j$ is a function of the integer sequence $j$. These shift operators obey the summation by parts
\begin{equation}
\sum_{j=-\infty}^{\infty}f_{j}\delta g_{j}=\sum_{j=-\infty}^{\infty}(\delta^{*}f_{j})g_{j},~~\ 
\sum_{j=-\infty}^{\infty}(\delta f_{j})g_{j}=\sum_{j=-\infty}^{\infty}f_{j}\delta^{*}g_{j}.
\label{summation_by_parts}
\end{equation}
If the lattice system has a boundary, the summation by parts yields a boundary term 
\begin{equation}
\sum_{j=1}^{\infty}f_{j}\delta g_{j}=\sum_{j=2}^{\infty}(\delta^{*}f_{j})g_{j}=\sum_{j=1}^{\infty}(\delta^{*}f_{j})g_{j}-f_{0}g_{1},~~\ 
\sum_{j=1}^{\infty}(\delta f_{j})g_{j}=\sum_{j=2}^{\infty}f_{j}\delta^{*}g_{j}=\sum_{j=1}^{\infty}f_{j}\delta^{*}g_{j}-f_{1}g_{0}.
\label{boundary_summation_by_parts}
\end{equation}
Using these properties, we can derive the edge Hamiltonians of the lattice system and their edge states. 

We next employ one-dimensional (1D) extended Su-Schrieffer-Heeger (SSH) model in the presence of the long-range hopping to illustrate the calculation of edge Hamiltonian. Its real-space Hamiltonian is
\begin{eqnarray}
H &=& \sum_j \left[ \hat{c}_j^\dagger \begin{pmatrix} 0 & \gamma \\ \gamma & 0 \end{pmatrix} \hat{c}_j + \hat{c}_j^\dagger \begin{pmatrix} 0 & t \\ \lambda & 0 \end{pmatrix} \hat{c}_{j+1} + \hat{c}_{j+1}^\dagger \begin{pmatrix} 0 & \lambda \\ t & 0 \end{pmatrix} \hat{c}_j \right] \nonumber\\
&=& \sum_j \hat{c}_j^\dagger \begin{pmatrix} 0 & \gamma + t {\delta} + \lambda \overleftarrow{{\delta}} \\ \gamma + \lambda {\delta} + t \overleftarrow{{\delta}} & 0 \end{pmatrix} \hat{c}_j \equiv \sum_j \hat{c}_j^\dagger \overleftrightarrow{\mathcal{H}} \hat{c}_j.
\label{generalized_ssh_model_hamiltonian}
\end{eqnarray}
where $\hat{c}_j \equiv (\hat{c}_{1j},\hat{c}_{2j})^\text{T}$,  $\hat{c}_{xj}$ ($x=1,2$) is the annihilation operator of the fermion in $x$th sublattice of the $j$th unit cell, $\gamma$ is the intracell hopping rate, $\lambda$ is the nearest-neighbor intercell hopping, and $t$ is the next-nearest-neighbor intercell hopping. Here $\overleftarrow{{\delta}}$ is an operator acting on the left shift. For an infinite system, Eq. \eqref{generalized_ssh_model_hamiltonian} is rewritten as
\begin{equation}
\begin{aligned}
H = \sum_{j=-\infty}^{\infty}\hat{c}_{j}^{\dagger} \overleftrightarrow{\mathcal{H}} \hat{c}_j=
\sum_{j=-\infty}^{\infty}\hat{c}_{j}^{\dagger}\hat{\mathcal{H}}\hat{c}_{j},
\end{aligned}
\end{equation}
where $\hat{\mathcal{H}}$ is defined by
\begin{equation}
\begin{aligned}
\hat{\mathcal{H}} = 
\begin{pmatrix}
0 & \gamma + \lambda {\delta}^* + t {\delta} \\
\gamma + \lambda {\delta} + t {\delta}^* & 0
\end{pmatrix}
\equiv \mathcal{K} {\delta}^* + \mathcal{K}^\dagger {\delta} + \mathcal{V}.
\end{aligned}
\end{equation}
Here $\hat{\mathcal{H}}$ operates only on the right-hand side operators, making it suitable for sovling the Schr\"{o}dinger eigen equation. The hat on $\hat{\mathcal{H}}$ emphasizes that it is not a simple matrix but rather a matrix operator. For a semi-infinite system, the Hamiltonian is given by
\begin{equation}
H=\sum_{j=1}^{\infty}\hat{c}_{j}^{\dagger} \overleftrightarrow{\mathcal{H}} \hat{c}_j=
\sum_{j=1}^\infty \hat{c}_j^\dagger\hat{\mathcal{H}}\hat{c}_j-\hat{c}_1\mathcal{K}\hat{c}_0,
\label{boundary_Hamiltonian}
\end{equation}
where the last term represents the excluded hopping interaction between the first unit cell and the $0$-th unit cell. 

On the other hand, we must consider the Hermiticity of Hamiltonian operator. For the wave functions of $\psi_j$ and $\phi_j$, we have 
\begin{equation}
\begin{aligned}
\langle \phi | \hat{\mathcal{H}} \psi \rangle 
& \equiv \sum_{j=1}^{\infty} \phi_j^\dagger \hat{\mathcal{H}} \psi_j 
= \sum_{j=1}^{\infty} \phi_j^\dagger (\mathcal{K} {\delta}^* + \mathcal{K}^\dagger {\delta} + \mathcal{V}) \psi_j\\
&= \sum_{j=1}^{\infty} \phi_j^\dagger (\mathcal{K} \overleftarrow{{\delta}} + \mathcal{K}^\dagger \overleftarrow{{\delta}^*} + \mathcal{V}) \psi_j - \phi_0^\dagger \mathcal{K}^\dagger \psi_1 + \phi_1^\dagger \mathcal{K} \psi_0\\
&\equiv \langle \hat{\mathcal{H}} \phi | \psi \rangle - \phi_0^\dagger \mathcal{K}^\dagger \psi_1 + \phi_1^\dagger \mathcal{K} \psi_0.
\end{aligned}
\label{Hermiticity_of_the_system's_Hamiltonian_operator}
\end{equation}
And then, the Hamiltonian can only be Hermitian if it satisfies the following equation:
\begin{equation}
\phi_0^\dagger \mathcal{K}^\dagger \psi_1 = 0, \quad \phi_1^\dagger \mathcal{K} \psi_0 = 0.
\label{Hermiticity_conditions}
\end{equation}
Namely, $\mathcal{K}$ must obey the condition
\begin{equation}
\phi_{m1}^\dagger\mathcal{K}\psi_{0n}=0
\label{Hermiticity_conditions——2}
\end{equation}
for all $m, n$. The $\psi_{jn}$($\phi_{mj}$) form a complete orthonormal set of functions $\psi_{j}$($\phi_{j}$). Among them, $\psi_{0n}$ is the edge states.

\subsubsection{b. Edge Hamiltonians of extended kagome lattice}

We first apply the above method to solve the edge states of the breathing kagome lattice. As the system only includes a nearest-neighbor hopping, the corresponding Hamiltonian is written as
\begin{equation}
\begin{split}
\hat{\cal H} =-\left( 
\begin{array}{ccc}
0 & t_{a}+t_{b}\delta^{*}_{\beta} & t_{a}+t_{b}\delta^{*}_{\alpha} \\ 
t_{a}+t_{b}\delta_{\beta} & 0 & t_{a}+t_{b}\delta^{*}_{\alpha}\delta_{\beta} \\ 
t_{a}+t_{b}\delta_{\alpha} & t_{a}+t_{b}\delta_{\alpha}\delta^{*}_{\beta} & 0%
\end{array}%
\right),
\end{split}
\label{NN_Kagome_Hamiltonian}
\end{equation}
where the $\mathcal{K}$ matrices for $\alpha$- and $\beta$-directions read
\begin{equation}
\begin{split}
\mathcal{K}_{\beta} =-t_b\left( 
\begin{array}{ccc}
0 & 1 & 0 \\ 
0 & 0 & 0 \\ 
0 & \delta_{\alpha} & 0%
\end{array}%
\right),~\mathcal{K}_{\alpha} =-t_b\left( 
\begin{array}{ccc}
0 & 0 & 1 \\ 
0 & 0 & \delta_{\beta} \\ 
0 & 0 & 0%
\end{array}%
\right) .
\end{split}
\label{NN_Kagome_K_matrices}
\end{equation}
The $\alpha$- and $\beta$-directions are along any two edges of the Kagome model, respectively. Note that $\delta_{\beta}$ in $\mathcal{K}_{\alpha}$ should be replaced by $e^{ik_{\beta}}$ when solving the edge states. We hereby consider the edge states in one direction and obtain
\begin{equation}
\mathcal{K}_{\alpha}\psi_{0n}=0.
\label{kagome_Hermiticity_conditions}
\end{equation}
Although each $\mathcal{K}$ matrix contains two non-zero elements, they are arranged vertically. Consequently, the edge states are easily given by
\begin{equation}
\psi_{0n}=
\begin{pmatrix}
\chi_n \\
0
\end{pmatrix},
\label{NN_Kagome_edge_states}
\end{equation}
where $\chi_n$ is a two-component vector. We can examine that this state satisfies $\mathcal{K}_{\alpha}\psi_{0n}=0$. Furthermore, we assume a Bloch wave function of the edge state as $\psi_{j_1n}=\psi_{0n}e^{iK_{\alpha,n}j_1}$. And then, the eigenvalue equation becomes
\begin{equation}
-
\begin{pmatrix}
0 & t_a+t_b e^{-ik_\beta} & t_a+t_b e^{-iK_{\alpha,n}} \\
t_a+t_b e^{ik_\beta} &0 & t_a+t_b e^{-iK_{\alpha,n}}e^{ik_\beta} \\
t_a+t_b e^{iK_{\alpha,n}} & t_a+t_b e^{iK_{\alpha,n}}e^{-ik_\beta} &0 
\end{pmatrix}
\begin{pmatrix}
\chi_n \\
0
\end{pmatrix}=\varepsilon_n
\begin{pmatrix}
\chi_n \\
0
\end{pmatrix}.
\label{NN_Kagome_edge_states_eigenvalue_equation}
\end{equation}
The $2\times2$ matrix in the upper left corner of the matrix gives ${\cal{H}}_{edge}$, i.e.,
\begin{equation}
{\cal{H}}_{edge}(k_\beta)=-
\begin{pmatrix}
 0& t_a+t_b e^{-ik_\beta}  \\
t_a+t_b e^{ik_\beta} &0 \\
\end{pmatrix}.
\label{NN_Kagome_H_edge}
\end{equation}
It is clear that this edge Hamiltonian is similar to a SSH model along $\beta$-direction. In $\alpha$-direction, the similar results are also obtained as follows:
\begin{equation}
{\cal{H}}_{edge}(k_\alpha)=-
\begin{pmatrix}
 0& t_a+t_b e^{-ik_\alpha}  \\
t_a+t_b e^{ik_\alpha} & 0\\
\end{pmatrix}.
\label{NN_Kagome_H_edge_a}
\end{equation}

Similarly, we can solve the edge states for the breathing kagome lattice with long-range hoppings. For convenience, we hereby consider the case shown in main text, i.e. $t_2\neq 0$ and $t_{m>2}=0$. The corresponding Hamiltonian of the system is given by
\begin{equation}
\begin{split}
\hat{\cal H}_{t_2} =&-\left( 
\begin{array}{ccc}
0 & t_{a}+t_{b}\delta^{*}_{\beta}+t_2{\delta^{*}_{\beta}}^2 & t_{a}+t_{b}\delta^{*}_{\alpha}+t_2{\delta^{*}_{\alpha}}^2 \\ 
t_{a}+t_{b}\delta_{\beta}+t_2{\delta_{\beta}}^2 & 0 & t_{a}+t_{b}\delta^{*}_{\alpha}\delta_{\beta}+t_2{\delta^{*}_{\alpha}}^2{\delta_{\beta}}^2 \\ 
t_{a}+t_{b}\delta_{\alpha}+t_2{\delta_{\alpha}}^2 & t_{a}+t_{b}\delta_{\alpha}\delta^{*}_{\beta}+t_2{\delta_{\alpha}}^2{\delta^{*}_{\beta}}^2 & 0%
\end{array}%
\right) ,\\
\end{split}
\label{eq:t2_Hamiltonian}
\end{equation}
of which the $\mathcal{K}$ and $\mathcal{L}$ matrices for $\alpha$- and $\beta$-directions read as
\begin{equation}
\begin{split}
\mathcal{K}_\alpha =-t_b\left( 
\begin{array}{ccc}
0 & 0 & 1 \\ 
0 & 0 & \delta_{\beta} \\ 
0 & 0 & 0%
\end{array}%
\right) ,
\mathcal{K}_\beta =-t_b\left( 
\begin{array}{ccc}
0 & 1 & 0 \\ 
0 & 0 & 0 \\ 
0 & \delta_{\alpha} & 0%
\end{array}%
\right) ,
\mathcal{L}_\alpha =-t_2\left( 
\begin{array}{ccc}
0 & 0 & 1 \\ 
0 & 0 & {\delta_{\beta}}^2 \\ 
0 & 0 & 0%
\end{array}%
\right) ,
\mathcal{L}_\beta =-t_2\left( 
\begin{array}{ccc}
0 & 1 & 0 \\ 
0 & 0 & 0 \\ 
0 & {\delta_{\alpha}}^2 & 0%
\end{array}%
\right) .
\end{split}
\label{t2_KL_matrices}
\end{equation}
The $\mathcal{L}$ matrix is a long-range interaction matrix. Further, we seek appropriate values for $\psi_{0n}$ and $\psi'_{0n}$, so that the conditions  
\begin{equation}
\mathcal{K}_{\alpha}\psi_{0n}=0,\ 
\mathcal{L}_{\alpha}\psi'_{0n}=0
\label{t2_kagome_Hermiticity_conditions}
\end{equation}
are satisfied. As the operators $\mathcal{K}_{\alpha}$ and $\mathcal{L}_{\alpha}$ commute and anticommute with each other, they share the same eigenvalues and can possess a common complete set of eigenvectors, 
\begin{equation}
\psi_{0n}=\psi'_{0n}=\begin{pmatrix}
\chi_n \\
0
\end{pmatrix}.
\label{t2_kagome_psi}
\end{equation}
Finally, we write down the eigenvalue equation,
\begin{equation}
-
\begin{pmatrix}
0 & t_a+t_b e^{-ik_\beta}+t_2 e^{-i2k_\beta}& h_{13} \\
t_a+t_b e^{ik_\beta}+t_2 e^{-i2k_\beta} & 0& h_{23} \\
h^{*}_{13} & h^{*}_{23} & 0
\end{pmatrix}
\begin{pmatrix}
\chi_n \\
0
\end{pmatrix}=\varepsilon_n
\begin{pmatrix}
\chi_n \\
0
\end{pmatrix}.
\label{t2_Kagome_edge_states_eigenvalue_equation}
\end{equation}
Here, $h_{13}=t_a+t_b e^{-iK_{\alpha,n}}+t_2 e^{-i2K_{\alpha,n}}$ and $h_{23}=t_a+t_b e^{-iK_{\alpha,n}}e^{ik_\beta}+t_2 e^{-i2K_{\alpha,n}}e^{i2k_\beta}$. 
Similarly, the $2\times2$ matrix in the upper left corner of the matrix gives the edge Hamiltonian
\begin{equation}
{\cal{H}}_{t_2,edge}(k_\beta)=-
\begin{pmatrix}
 0& t_a+t_b e^{-ik_\beta}+t_2 e^{-i2k_\beta}  \\
t_a+t_b e^{ik_\beta}+t_2 e^{-i2k_\beta} & 0\\
\end{pmatrix},
\label{t2_Kagome_H_edge}
\end{equation}
It is clear that ${\cal{H}}_{t_2,edge}(k_\beta)$ is similar to an SSH model with long-range hopping. More generally, the edge states of the extended kagome lattice are characterized by an extended SSH model,
\begin{equation}
{\cal{H}}_{t_m,edge}(k_\beta)=-
\begin{pmatrix}
 0& t_a+t_b e^{-ik_\beta}  \\
t_a+t_b e^{ik_\beta}& 0\\
\end{pmatrix}
-\sum_{m=2}^{M}t_m
\begin{pmatrix}
 0&  e^{-imk_\beta}  \\
 e^{-imk_\beta} & 0\\
\end{pmatrix}.
\label{tn_Kagome_H_edge}
\end{equation}
In $\alpha$-direction, the similar results are also obtained as follows:
\begin{equation}
{\cal{H}}_{t_m,edge}(k_\alpha)=-
\begin{pmatrix}
 0& t_a+t_b e^{-ik_\alpha}  \\
t_a+t_b e^{ik_\alpha}& 0\\
\end{pmatrix}
-\sum_{m=2}^{M}t_m
\begin{pmatrix}
 0&  e^{-imk_\alpha}  \\
 e^{-imk_\alpha} & 0\\
\end{pmatrix}.
\label{tn_Kagome_H_edge}
\end{equation}

\subsubsection{c. 2D topological characterization theory}
 The above edge Hamiltonian can be generally written as 
\begin{equation}
{\cal H}_{edge}(k_\beta)=d'_{\beta}(k_\beta)\sigma_x+d_{\beta}(k_\beta)\sigma_y.
\label{x_SSH_model}
\end{equation}
Then, we obtain $d'_{\beta}=-t_a-t_b\cos k_\beta-t_m\sum^M_{m=2}\cos (mk_\beta)$ and $d_{\beta}=-t_b\sin k_\beta-t_m\sum^M_{m=2}\sin (mk_\beta)$. This implies that each $1$D edge of this extended kagome lattice can be described by ${\cal H}_{edge}(k_j)$, where $k_j=\mathbf{k}\cdot{\mathbf{a}_j}$ and $j=1,2,3$. Hence, this 2D extended kagome lattice have the second-order topology and can host corner states, which is characterized by
\begin{equation}
W=W_{k_\beta}W_{k_\alpha}.
\end{equation}
Here $W_{k_\beta}$ ($W_{k_\alpha}$) defines a winding number of $1$D edge Hamiltonian ${\cal H}_{edge}(k_\beta)$ [${\cal H}_{edge}(k_\alpha)$] along the $k_\beta$ ($k_\alpha$) directions. Note $\alpha$ and $\beta$ can take an arbitrary element of the set $\{1,2,3\}$ and $\alpha\neq \beta$. Remarkably, $W_{k_\alpha}$  and $W_{k_\beta}$ can be determined by the polarized topological charges~\cite{s-PhysRevB.110.L201117}. For example of $W_{k_\beta}$, we use the convention that $d'_{\beta}(k_\beta)$ denotes the band dispersion, while $d_{\beta}(k_\beta)$ denotes the pseudospin-orbit coupling. Then, a topological charge is located at $d'_{\beta}(\varrho_l)=0$ and can be defined as 
\begin{equation}
{\cal C}^{k_\beta}_l=\text{sgn}\Big[\frac{\partial d'_{\beta}(\varrho_l)}{\partial{k_\beta}}\Big].
\label{k_beta_charges}
\end{equation}
The corresponding charge polarization is given
\begin{equation}
{\cal P}^{k_\beta}_l=\text{sgn}\left[d_{\beta}(\varrho_l)\right]. 
\label{xpolarization}
\end{equation}
Then, we can obtain a polarized topological charge $\mathcal{S}^{k_\beta}_l={\cal C}^{k_\beta}_l{\cal P}^{k_\beta}_l$. One can analytically obtain 
\begin{equation}
W_{k_\beta}=-\frac{1}{2}\text{sgn}\left[\mathcal{N}_{+}^{k_\beta}\mathcal{N}_{-}^{k_\beta}\right]\sum_{l=1}^{\mathcal{N}^{k_\beta}}\mathcal{S}^{k_\beta}_l,
\label{k_beta_topo}
\end{equation}
where $\mathcal{N}_{+}^{k_\beta}$($\mathcal{N}_{-}^{k_\beta}$) is the number of nodal points with $d_{\beta}(k_\beta)=0$ in region of $d'_{\beta}(k_\beta)>0$ [$d'_{\beta}(k_\beta)<0$]. Here $\mathcal{N}^{k_\beta}$ is the number of polarized topological charges. In $k_\alpha$ direction, the similar results are also obtained as follows:
\begin{equation}
W_{k_\alpha}=-\frac{1}{2}\text{sgn}\left[\mathcal{N}_{+}^{k_\alpha}\mathcal{N}_{-}^{k_\alpha}\right]\sum_{m=1}^{\mathcal{N}^{k_\alpha}}{\cal S}^{k_\alpha}_{m},
\label{k_alpha_topo}
\end{equation}
The similar results can be obtained for $W_{k_\alpha}$. Finally, the second-order topological phase of the extended kagome lattice can be characterized by 
\begin{equation}
P=\frac{1}{4}\text{sgn}\left[\mathcal{N}_{+}^{k_\beta}\mathcal{N}_{-}^{k_\beta}\mathcal{N}_{+}^{k_\alpha}\mathcal{N}_{-}^{k_\alpha}\right]
\sum_{l=1}^{\mathcal{N}^{k_\beta}}{\cal S}^{k_\beta}_{l}
\sum_{m=1}^{\mathcal{N}^{k_\alpha}}{\cal S}^{k_\alpha}_{m},
\label{topo_characterization}
\end{equation}
It should be noted that Eq.~\eqref{topo_characterization} can be rewritten as $P=P_1P_2$, with 
\begin{equation}
P_1=\text{sgn}\left[\mathcal{N}_{+}^{k_\beta}\mathcal{N}_{-}^{k_\beta}\mathcal{N}_{+}^{k_\alpha}\mathcal{N}_{-}^{k_\alpha}\right],~~P_2=\frac{1}{4}\sum_{l=1}^{\mathcal{N}^{k_\beta}}{\cal S}^{k_\beta}_{l}
\sum_{m=1}^{\mathcal{N}^{k_\alpha}}{\cal S}^{k_\alpha}_{m}.
\end{equation}
When we further define $\boldsymbol{\varrho}\equiv\{\mathbf{k}\in\text{BZ}|d_\alpha(\mathbf{k})=d_\beta(\mathbf{k})=d_\gamma(\mathbf{k})=0\}$, $P_1$ can be combined as $P_1=\text{sgn}(\mathcal{N}_+\mathcal{N}_-)$, where $\mathcal{N}_+$ and $\mathcal{N}_-$ are the number of $\boldsymbol{\varrho}$ points in the regions of ${d'_\gamma}(\boldsymbol{\varrho})>0$ and ${d'_\gamma}(\boldsymbol{\varrho})<0$, respectively. Similarly, $P_2$ can also be combined as $P_2=\sum_{n}\mathcal{C}_n\mathcal{P}_n$, where $\mathcal{C}_n=\text{sgn}[\frac{\partial d^{\prime}_{\alpha}(\boldsymbol{\vartheta})}{\partial k_{\alpha}}
\frac{\partial d^{\prime}_{\beta}(\boldsymbol{\vartheta})}{\partial k_{\beta}}]$
define the $n$th topological charge located at $\boldsymbol{\vartheta}\equiv\{\mathbf{k}\in\text{BZ}|d'_\alpha(\mathbf{k})=d'_\beta(\mathbf{k})=0\}$. Its polarization is defined as 
${\cal P}_n=\text{sgn}[d_{\alpha}(\boldsymbol{\vartheta})d_{\beta}(\boldsymbol{\vartheta})]$. Finally, the topological invariant is given by
\begin{equation}
\begin{split}
{P}=\frac{1}{4}\text{sgn}(\mathcal{N}_+\mathcal{N}_-)
\sum_{n}\mathcal{C}_n\mathcal{P}_n=\frac{1}{4}\text{sgn}(\mathcal{N}_+\mathcal{N}_-)
\sum_{n}\mathcal{S}_n,\\
\end{split}
\label{topo_characterization_last}
\end{equation}
which exactly characterize the zero-energy corner states in the extended kagome lattice. 

\subsubsection{d. Other cases with more long-range hopping terms}

When adding the second nearest neighbor hopping $t_c$, the corresponding Bloch Hamiltonian is written as
\begin{equation}
\begin{split}
\hat{\cal H} =-\left( 
\begin{array}{ccc}
0 & t_{a}+t_{b}\delta^{*}_{\beta} & t_{a}+t_{b}\delta^{*}_{\alpha} \\ 
t_{a}+t_{b}\delta_{\beta} & 0 & t_{a}+t_{b}\delta^{*}_{\alpha}\delta_{\beta} \\ 
t_{a}+t_{b}\delta_{\alpha} & t_{a}+t_{b}\delta_{\alpha}\delta^{*}_{\beta} & 0%
\end{array}%
\right)+\hat{\cal H}_c,
\end{split}
\label{NNc_Kagome_Hamiltonian}
\end{equation}
Here, $\hat{\cal H}_c$ denotes the Hamiltonian for the next-nearest-neighbor hopping and reads
\begin{equation}
\begin{split}
\hat{\cal H}_c =-t_c\left( 
\begin{array}{ccc}
0 & \delta_{\alpha}\delta^{*}_{\beta}+\delta^{*}_{\alpha} & \delta^{*}_{\beta}+\delta^{*}_{\alpha}\delta_{\beta} \\ 
\delta^{*}_{\alpha}\delta_{\beta}+\delta_{\alpha} & 0 & \delta^{*}_{\alpha}+\delta_{\beta} \\ 
\delta_{\beta}+\delta_{\alpha}\delta^{*}_{\beta} & \delta_{\alpha}+\delta^{*}_{\beta} & 0%
\end{array}%
\right).
\end{split}
\label{NNc_Kagome_Hamiltonian}
\end{equation}
At this point, the $\mathcal{K}$ matrices becomes complicated, i.e.,
\begin{equation}
\begin{split}
&\mathcal{K}_{\beta} =-t_b\left( 
\begin{array}{ccc}
0 & 1 & 0 \\ 
0 & 0 & 0 \\ 
0 & \delta_{\alpha} & 0%
\end{array}%
\right)-t_c\left( 
\begin{array}{ccc}
0 & \delta_{\alpha} & 1 \\ 
0 & 0 & 0 \\ 
\delta_{\alpha} & 1 & 0%
\end{array}\right),\\
&\mathcal{K}_{\alpha} =-t_b\left( 
\begin{array}{ccc}
0 & 0 & 1 \\ 
0 & 0 & \delta_{\beta} \\ 
0 & 0 & 0%
\end{array}%
\right)-t_c\left( 
\begin{array}{ccc}
0 & 1 & \delta_{\beta} \\ 
\delta_{\beta} & 0 & 1 \\ 
0 & 1 & 0%
\end{array}\right) .
\end{split}
\label{NN_Kagome_K_matrices}
\end{equation}
It is very difficult to find a suitable $\psi_{0n}$ that satisfies this condition,
\begin{equation}
\mathcal{K}_{\alpha}\psi_{0n}=0.
\label{kagome_Hermiticity_conditions}
\end{equation}
Then, the Hamiltonian can no longer be projected onto the boundary. As a result, our topological characterization method becomes invalid.

\begin{figure}[h]
\centering
\includegraphics[width=1\columnwidth,trim=0cm 0 0cm 0cm,clip=false]{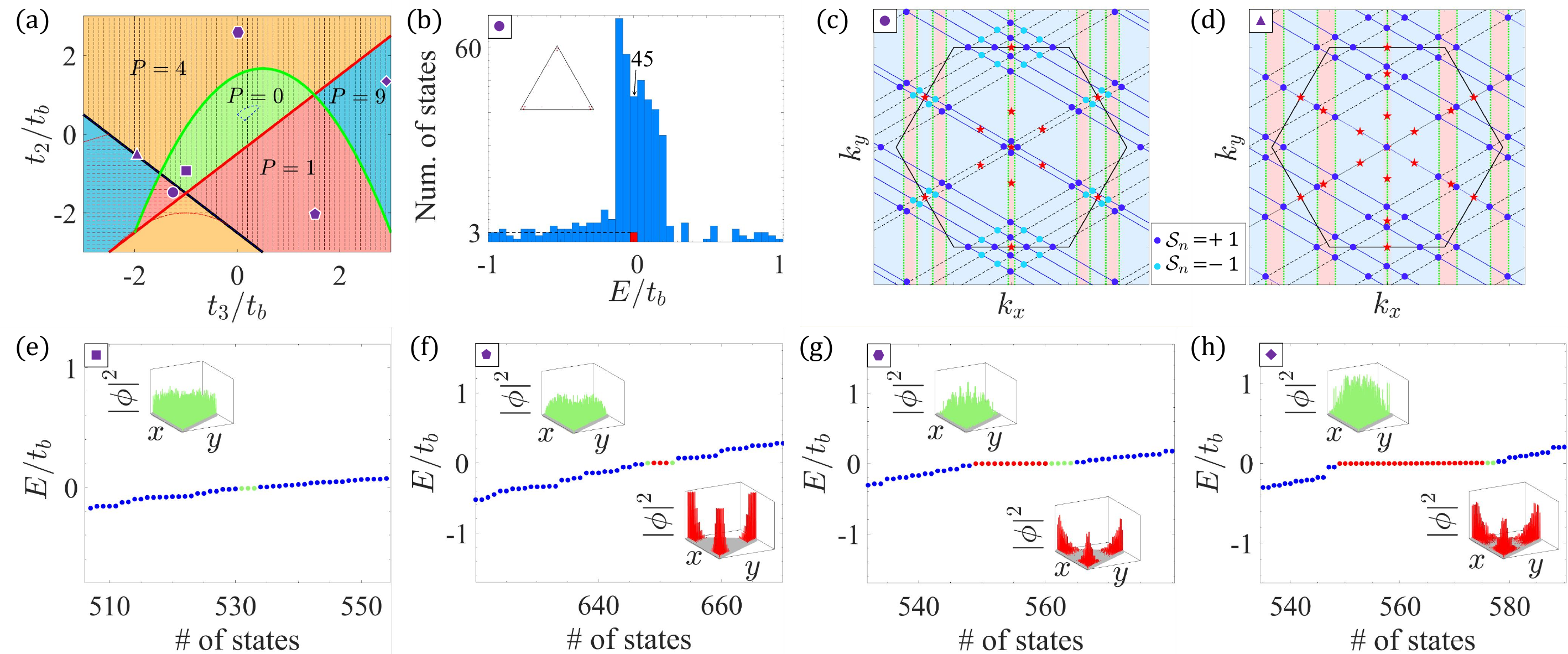}
\caption{(a) Topological phase diagram determined by $P$. The shadow black vertical dashed (red horizontal dashed) lines mark the regions where zero-energy is located at the upper (lower) of bulk energy gap. Here the parameters are $t_a=1.5t_b,\ t_{m>3}=0\ $.
(b) Statistics diagram of the eigenstates of the Hamiltonian under the $x$- and $y$-direction open boundary condition, where there are 3 topological zero-energy corner states and characterized by $P=1$. The inset show the distribution of zero-energy states. Here the parameters are $t_2=-1.5t_b,\ t_3=-1.1t_b$.
(c) Topological characterization for $P=1$, $t_2=-1.5t_b$, and $t_3=-1.1t_b$. 
(d) Type-III TPT for $t_2=-0.5$ and $t_3=-2$. 
(e)-(h) Distribution zero-energy states in the continuum. 
The zero-energy topological corner states are labeled in red and the zero-energy bulk states are labeled in green, as well as their probability distributions calculated in the inset.
The parameters are 
 $t_2=-1t_b,\ t_3=-1t_b$ for (e),
 $t_2=-2t_b,\ t_3=1.5t_b$ for (f),
 $t_2=2.5t_b,\ t_3=0t_b$ for (g), and
$t_2=1.3t_b,\ t_3=2.9t_b$ for (h). 
}
\label{fig:sm2}
\end{figure}

\subsection{II. More numerical results of extended kagome model}\label{two}
We provide more parameter cases of extended kagome lattice, where the system includes longer-range hopping $t_3$. The corresponding Hamiltonian reads 
\begin{equation}
\begin{split}
H(\mathbf{k}) =&-\left( 
\begin{array}{ccc}
0 & h_{12} & h_{13} \\ 
h_{12}^{\ast } & 0 & h_{23} \\ 
h_{13}^{\ast } & h_{23}^{\ast } & 0%
\end{array}%
\right),
\end{split}
\label{eq:Hamiltonian2}
\end{equation}
with $h_{12}=t_{a}+t_{b}e^{-i\mathbf{k\cdot a}_{3}}+\sum^M_{m=2}t_{m}e^{-im\mathbf{k\cdot a}_{3}}$, $h_{13}=t_{a}+t_{b}e^{-i\mathbf{k\cdot a}_{2}}+\sum^M_{m=2}t_{m}e^{-im\mathbf{k\cdot a}_{2}}$, and $h_{23}=t_{a}+t_{b}e^{-i\mathbf{k\cdot a}_{1}}+\sum^M_{m=2}t_{m}e^{-im\mathbf{k\cdot a}_{1}}$. Using the theory of topological characterization, we obtain the phase diagram in Fig.~{\ref{fig:sm2}}(a). It is observed that the topologically protected BICs are present in a broad parameter region. In the case of $t_2=-1.5t_b$ and $t_3=-1.1t_b$, the system hosts $45$ zero-energy states in
the continuum [see Fig.~{\ref{fig:sm2}}(b)], but only three of them are localized at the corners. Figure~{\ref{fig:sm2}}(c) shows the distribution of the topological charges and their polarizations. Its $\boldsymbol{\varrho}$ momentum points (see red stars) are located in both positive and negative region of ${d'_\gamma}(\boldsymbol{\varrho})$, which gives $P_1=1$.  Meanwhile, there are $20$ positive and $16$ negative polarized topological charges, which gives $P_2=\sum_n\mathcal{C}_n/4=1$. Thus, the corresponding topological invariant is $P=P_1P_2=1$. In addition, three types of topological phase transition (TPT) are also observed in Fig.~{\ref{fig:sm2}}(a), which are similar to the case of the main text. These TPTs are still obtained by our scheme. Figure~\ref{fig:sm2}(d) shows an example of type-III TPT. We take more parameters to confirm that the corner states in the continuum are captured by our characterization theory. In Figs.~{\ref{fig:sm2}}(e)-(h), we present the distribution of zero-energy states for $P=0$, $1$, $4$, and $9$, respectively. It is seen that $3P$ zero-energy states in the continuum are localized at the corners and the others are located in the bulk. These results further demonstrate the advantages of our topological characterization theory in the extended kagome lattice.

\end{document}